\documentstyle[prl,preprint,tighten,aps,floats]{revtex}
\newcommand{\tj}{$t$-$J$ }
\newcommand{\bq}{{\bf q}}
\newcommand{\bk}{{\bf k}}
\newcommand{\bkp}{{\bf k'}}
\newcommand{\ybco}{YBa$_2$Cu$_3$O$_{6+\delta}$}
\newcommand{\cusp}{\Delta (d\mu^e \!/ dT)}

\begin{document}

\title {Temperature dependence of the electron chemical potential
  \protect\\ in \ybco}

\author {A. V. Dotsenko and O. P. Sushkov$^{*}$}
\address {School of Physics, The University of New South Wales,
                Sydney 2052, Australia}

\maketitle

\begin{abstract}
We study the behaviour of the electronic chemical potential in \ybco\
  near the superconducting transition
  using the spin-wave exchange theory of pairing.
We find that the experimental value of the jump
  in the temperature derivative of the chemical potential
  is inconsistent with the theoretical value calculated
  under the assumption of constant hole concentration.
It is suggested that charge transfer between chains and planes
  takes place and corresponding calculations are done.
\end{abstract}


The temperature dependence of the chemical potential
  near the superconducting transition has attracted some attention 
  recently~\cite{Rob2,Ric8,Ran9,VDM1,Kh,VDM2,VDM3,VDMexp}.
It has been shown in several models that there is a jump
  in the temperature derivative of the chemical potential.
In Ref.~\cite{Rob2},
  in a simple BCS model with quadratic dispersion and $s$-wave pairing,
  the relationship between
  the chemical potential in the superconducting state $\mu_s$
  and the chemical potential in the normal state $\mu_n$ was found to be
\begin{equation}  \label{mus}
  \mu_s = \mu_n - {{\mit\Delta}^2 \over 4 \mu_n},
\end{equation}
  where ${\mit\Delta}$ is the gap at the Fermi surface.
Near the critical point
  \mbox{\( {\mit\Delta} \propto \sqrt {T_c - T} \)},
  and from Eq.~(\ref{mus}) it follows that
  the temperature dependence of the chemical potential is not smooth.
This effect has been observed experimentally~\cite{VDMexp}. 
Measurements of the work function of the \ybco\ compound with
  the critical temperature of 90\,K gave
\begin{equation} \label{exp}
  {d(\mu_n^e - \mu_s^e) \over dT}  {\biggr |} _{T = T_c}
  = 0.12 \pm 0.02.
\end{equation}
(Throughout the paper, the Boltzmann constant $k_B = 1$.)

The purpose of the paper is to present a quantitative study of
  this effect specifically for \ybco.
Using the magnetic fluctuation exchange theory
  of high temperature superconductivity,
  we find a value of $\cusp$
  much larger than the experimental value~(\ref{exp})
  and we show that the discrepancy may be explained
  by taking into account the presence of chains.

\bigskip 
Before calculations, let us discuss various contributions into
  the temperature dependence of the chemical potential.
The quantity measured in the experiment~\cite{VDMexp}
  is the single electron work function.
To analyse it
  we must consider the single particle Green's function.
The mean field approximation corresponds to the account of
  the leading order anomalous self-energy (Fig.~1)
  in the Dyson equations
  for the normal and anomalous Green's functions~\cite{Lif}.
With a normalisation condition, it gives an effect as in Eq.~(\ref{mus}). 
In the next order,
  there is the self-energy contribution presented in Fig.~2, which
  is due to modification of the spin-wave (wavy line) at the transition point.
Analysis of this diagram is very complicated.
At zero temperature, this problem was recently studied in~\cite{Sus95}. 
We plan to analyse the problem more generally in the future.
In the present work we neglect the contribution of the diagram in Fig.~2
  {\it i.e.\/}~we stay within the mean field approximation.

Another aspect to consider
  is the external influence on the electronic subsystem.
Equation~(\ref{mus}) corresponds to
  the usual BCS mean field theory with constant electronic density.
The experiment corresponds to constant external pressure,
  however due to interaction with the lattice,
  the electronic subsystem is
  neither at constant density nor at constant pressure.
[It appropriate to note here that,
  at constant electronic pressure,
  according to Landau theory of second order phase transitions
  there is no discontinuity in $d\mu/dT$ (see discussion in Ref.~\cite{VDM3}).]
We thus have to understand first which theoretical value
  should be compared with the experiment.

There is a rearrangement of the lattice
  under the superconducting transition.
Its influence on the single electron
  Green's function is described by the diagram in Fig.~3,
  with the dash line representing phonons.
Since there are two anomalous Green's functions in the loop
  this contribution is proportional to ${\mit\Delta}^2$.
However, it can be neglected
  if the interaction with phonons is weak and the lattice is hard,
We therefore take the electron desity to be constant.

\bigskip 
Let us now outline the mechanism of pairing
  induced by spin-wave exchange in the \tj model~\cite{Flam94,Bel95}.
The \tj model is defined by the Hamiltonian 
\begin{equation} \label{H}
  H = - t \!\! \sum_{<nm>\sigma} 
          ( d_{n\sigma}^{\dag} d_{m\sigma}^{} + \text{H.c.} )
    +  J \!\! \sum_{<nm>}  {\bf S}_n \cdot {\bf S}_m,
\end{equation}
  where $d_{n\sigma}^{\dag}$ is the creation operator of a hole with spin
  $\sigma$ ($\sigma= \uparrow, \downarrow$)
  at site $n$ of a two-dimensional square lattice and
  $<nm>$ are nearest-neighbour sites.
The $d_{n\sigma}^{\dag}$ operators act in the Hilbert space
  with no double electron occupancy.
The spin operator is
  \mbox{\( {\bf S}_n = \case{1}{2} d_{n \alpha}^{\dag}
  ${\boldmath $\sigma$}$_{\alpha \beta} d_{n \beta}  \)}.
Below we set $J$ and the lattice spacing equal to unity.

At half-filling (one hole per site), the \tj model is equivalent to
  the Heisenberg antiferromagnetic model
  which has long-range order in the ground state.
We denote the wave function of this (undoped) ground state by $|0\rangle$
  and use it as the starting point for considering the doped state.
It is well known that doping destroys long-range antiferromagnetic order. 
However, we may use $|0\rangle$ and the corresponding set of excitations 
  as the basis in the problem with doping. 
The effective Hamiltonian for the \tj model
  as derived in~\cite{Cher4,Suhf,Kuch3} is
\begin{equation}  \label{Heff}
 H_{\text{eff}}
     =  \sum_{\bk\sigma}  \epsilon_\bk
                      h_{\bk\sigma}^{\dag}   h_{\bk\sigma}^{}
     +
         \sum_\bq   \omega_\bq
                     (   \alpha_\bq^{\dag}  \alpha_\bq^{}
                        + \beta_\bq^{\dag}  \beta_\bq^{}
                     )
      +  H_{\text{h,sw}}
      +  H_{\text{h,h}}.
\end{equation}
It is expressed in terms of the usual spin waves on antiferromagnetic
  background $\alpha_\bq$ and $\beta_\bq$ 
  and composite hole operators $h_{\bk\sigma}$
  ($\sigma = \pm 1/2$ is the pseudo-spin index).
The summations over $\bk$ and $\bq$ are
  over the Brillouin zone of one sublattice
  (defined by \mbox{$|k_x| + |k_y| \leq \pi$}).

The wave function of a composite hole can be represented as
  $\psi_{\bk\sigma} = h_{\bk\sigma}^{\dag} |0\rangle$.
At large~$t$ the operator $h_{\bk\sigma}^{\dag}$ has a complicated structure.
At $t = 3$, the weight of the bare hole in $\psi_{\bk\sigma}$ is about 25\,\%, 
  the weight of configurations ``bare hole + 1 magnon'' is  $\sim$\,50\,\%,
  and the rest are more complicated configurations.
The dressed hole is a normal fermion. 
For $t \lesssim 5$ the dispersion can be 
  approximated by the expression~\cite{Sus2}
\begin{equation}  \label{hdisp}
 \epsilon_\bk = \text{const}
                - \sqrt{0.43 + 4.6\, t^2 - 2.8\, t^2 \gamma_\bk^2}\,
                + \case{1}{4} \beta_2 (\cos k_x - \cos k_y)^2,
\end{equation}
  where $\gamma_\bk = \case{1}{2} (\cos k_x + \cos k_y)$.
We use the rigid band approach, {\it i.e.\/}\ we always use the dispersion
  Eq.~(\ref{hdisp}) which was derived for the undoped model.
The basis for this approach is that the results are not highly sensitive to
  the dispersion and there is some evidence, including experimental,
  that the changes in the dispersion caused by doping are moderate.
Near the band minima $\bk_0 = (\pm \pi/2, \pm\pi/2)$,
  the dispersion (\ref{hdisp}) can be 
  presented in the usual quadratic form
\begin{equation} \label{hdisp1}
  \epsilon_\bk = \case{1}{2} \beta_1^{} p_1^2 + \case{1}{2} \beta_2^{} p_2^2,
\end{equation}
  where $p_1$ ($p_2$) is the projection of $\bk - \bk_0$
  on the direction orthogonal (parallel)
  to the face of the magnetic Brillouin zone,
  and $\beta_1 \approx 0.65\, t$ at $t \gg 0.3$. 
The coefficient $\beta_2$ is small ($\beta_2 \ll \beta_1$)
  and the dispersion is almost degenerate
  along the edge of the magnetic Brillouin zone ($\gamma_\bk = 0$).
According to Refs.~\cite{Mart1,Liu2,puretj},
  $\beta_2 \approx 0.1\, t$ at $t \gtrsim 0.3$.
For the physical value of $t \approx 3$
  it means $\beta_2 \approx 0.3$.
Angle resolved photoemission experiments~\cite{arpes} indicate a smaller
  value of $\beta_2 \simeq 0.1$, which can be modelled by adding
  a small diagonal hopping term to the original \tj Hamiltonian.
While being crucial in some experiments (such as the Hall effect),
  when pairing is considered,
  the exact value of $\beta_2$ is not very important as long as it is small. 
In the present paper we use $\beta_2 = 0$ and $\beta_2 = 0.3$.

The spin-wave dispersion in (\ref{Heff}) is 
\begin{equation}  \label{swdisp}
 \omega_\bq = 2 \sqrt {1 - \gamma_\bq^2}.
\end{equation}
A certain approximation is made here since in the presence of mobile holes
  long range antiferromagnetic order is destroyed
  and a gap is formed in the spin-wave spectrum.
The used approximation assumes that magnetic order is preserved
  at the scales most relevant to pairing.
We checked that 
  the results are not dramatically affected if a small gap is introduced
  in~Eq.~(\ref{swdisp}).

The interaction of composite holes with spin waves
  is of the form~\cite{Suhf,Mart1,Liu2,Kan9}
\begin{eqnarray}  \label{hsw}
 H_{\text{h,sw}}
  &&  = \sum_{\bk,\bq} g_{\bk,\bq}
          \left(
                h_{\bk+\bq\downarrow}^{\dag} h_{\bk\uparrow}^{} \alpha_\bq^{}
              + h_{\bk+\bq\uparrow}^{\dag} h_{\bk\downarrow}^{} \beta_\bq^{}
              + \text{H.c.}
         \right),
 \nonumber\\
 g_{\bk,\bq}
  && = 2 f
        \left(  {2\over N}  \right)^{1/2}
        \left(
              \gamma_\bk U_\bq + \gamma_{\bk+\bq} V_\bq
        \right),
\end{eqnarray}
  where  $N$ is the total number of sites,
  \( U_\bq = \sqrt{ {1 \over \omega_\bq} + {1\over 2} } \)
    and
  \( V_\bq = - \text{sign}  (\gamma_\bq)
               \sqrt{ {1\over \omega_\bq } - {1\over 2} } \)
  are the parameters of the Bogolyubov transformation diagonalising
  the spin-wave Hamiltonian. 
At $t = 3$, the coupling constant $f$ in Eq.~(\ref{hsw})
  is $f = 1.8$~\cite{Suhf}.
It is strongly renormalised (by a factor of order $J/t$)
  compared to the coupling constant for bare holes
  but, importantly, the kinematic structure of the vertex
  remains the same (see~\cite{Suhf,Mart1,Liu2,Kan9}).

Finally, there is a contact hole-hole interaction $H_{\text{h,h}}$
  in the effective Hamiltonian (\ref{Heff}).
It is proportional to some function $A(t)$.
For small $t$ this function approaches the value $A(0) = -\case{1}{4}$,
  which reflects the well known hole-hole attraction
  induced by reduction of the number of missing antiferromagnetic links.
However $A(t)$ vanishes at $t \approx 3$~\cite{Cher4}.
We therefore neglect the contact interaction $H_{\text{h,h}}$
  and consider only the hole-spin-wave interaction $H_{\text{h,sw}}$.

Because of the nearly flat hole dispersion and extremely low Fermi energy
  (\mbox{$\epsilon_F \sim 100$\,K} at optimal doping),
  the spin wave velocity is much higher than the typical hole velocity.
It means that the spin-wave exchange interaction is
  essentially instantaneous and retardation can be ignored.
The wave function of the superconducting state is constructed
  in the usual BCS form involving dressed holes
\begin{equation} \label{psi}
 | \Psi \rangle = \prod_\bk
                  \left(
                         u_\bk + v_\bk
                           h_{\bk\uparrow}^{\dag} h_{-\bk\downarrow}^{\dag}
                  \right)
                  | 0 \rangle,
\end{equation}
  where $u_\bk^2 + v_\bk^2 = 1$ and $|0\rangle$ is the undoped state.
For the gap we then have the usual BCS equation
\begin{eqnarray}  \label{BCS}
&& {\mit\Delta}_\bk
      = - \sum_\bkp V_{\bk\bkp}
                   {{\mit\Delta}_\bkp \over 2E_\bkp}
                   \tanh {E_\bkp \over 2T},
\\
&&
  E_\bk = \sqrt{\xi_\bk^2 + {\mit\Delta}_\bk^2},
\qquad
  u_\bk v_\bk = {{\mit\Delta}_\bk \over 2 E_\bk},
\qquad
  \xi_\bk = \epsilon_\bk - \mu,
\nonumber
\end{eqnarray}
  with the chemical potential $\mu$ fixed by the hole concentration
\begin{equation} \label{del}
 x = \sum_\bk
           \left(
                  1 - {\xi_\bk  \over  E_\bk} \tanh {E_\bk \over 2T}
           \right).
\end{equation}
The interaction in Eq.~(\ref{BCS}) is
\begin{equation} \label{Vkk}
 V_{\bk \bkp} 
    = -2  {
                g_{-\bk, \bq} g_{-\bkp, \bq}
          \over
                - \omega_\bq - |\xi_\bk| - |\xi_\bkp|,
           }
\end{equation}
  where $\bq = \bk + \bk'$.
The diagrams representing the exchange of one spin wave are shown in Fig.~4.
The kinematic structure of the interaction is better seen
  if we use Eq.~(\ref{hsw}) and write
\begin{equation} \label{gg}
 g_{-\bk, \bq} g_{-\bkp, \bq}
      = { 4 f^2  \over N } \cdot
        {
            2 \gamma_\bk \gamma_\bkp
         -  (\gamma_\bk^2 + \gamma_\bkp^2)   \gamma_\bq 
        \over
            (1 - \gamma_\bq^2) ^{1/2}
        }.
\end{equation}
The structure of the interaction~(\ref{Vkk}) resulting from
  the hole-spin-wave vertex of Eq.~(\ref{gg})
  is an important difference of the used approach
  from the phenomenological theories of superconductivity~\cite{pines}
  which take the hole-spin-wave vertex as a constant
  without any $\bk$-dependence.
We believe that Eq.~(\ref{Vkk}) is more appropriate in the context of
  strongly correlated systems since it is derived from the microscopic
  Hamiltonian.

Details of an approximate analytical and exact numerical solutions
  of Eq.~(\ref{BCS}) are discussed in \cite{Flam94} and~\cite{Bel95}.
There is an infinite set of solutions with different symmetries but
  the strongest pairing is in the $d_{x^2-y^2}$-wave channel.

\bigskip  
The calculated chemical potential for hole concentration $x = 0.15$
  is presented in Fig.~5.
To study analytically the temperature dependence of the chemical potential,
  it is convenient to use partial derivatives of concentration.
For the jump in the temperature derivative of the chemical potential we have
\begin{equation} \label{pur}
 \left.  {d(\mu_n - \mu_s) \over dT}  \right| _{T = T_c}
   = {
         ( \partial x / \partial T   )_s
     \over
         ( \partial x / \partial \mu )_s
     }
   - {
         ( \partial x / \partial T )_n
     \over
         ( \partial x / \partial \mu )_n
     },
\end{equation}
  where indices $s$ and $n$ refer to derivatives being taken
  below and above the transition point, respectively.
In experiments, electron rather than hole chemical potential
  is measured and the appropriate quantity is therefore
  $\cusp = - \Delta (d\mu / dT)$.

The results of our calculations are presented in Table I.
\begin{table}[bt] 
\caption{
  Partial derivatives of hole concentration at the transition point.
  The hole concentration is $x = 0.15$. }
\begin{tabular}{cccccccc}
 $\beta_2$  & $\mu(0)$  & $\mu(T_c)$  &  $T_c$ &
     $ ( \partial x / \partial \mu )_n $   &
     $ ( \partial x / \partial \mu )_s $   &
     $ ( \partial x / \partial T   )_n $   &
     $ ( \partial x / \partial T   )_s $
\\ \tableline
0.0 & 0.006 & 0.010 &  0.042 & 1.97 & 2.45 & 0.92 & $-$0.55  \\
0.3 & 0.134 & 0.139 &  0.039 & 1.36 & 1.69 & 0.84 & $-$0.44  \\
\end{tabular} \end{table} 

Substituting data from Table I into Eq.~(\ref{pur}) or from Fig.~5, we find
      $\cusp = 0.69$ for $\beta_2 = 0  $
  and $\cusp = 0.88$ for $\beta_2 = 0.3$,
  which is more than 5 times larger than the experimental value~(\ref{exp}).
In the rest of the paper we examine
  how the presence of chains may affect this result.

\bigskip 
The chains in \ybco\ are known to be charge reservoirs and we can expect
  that the presence of a reservoir
  will reduce variations of the chemical potential
  (in the extreme case of an infinite reservoir
     the chemical potential is always constant).
The conserved quantity is the total hole concentration
  $\delta = 2x + x'$, where $x$ is the in-plane hole concentration,
  $x'$ is the hole concentration in the chains, and we have
  taken in account that there is one chain for two planes.
The expression for the jump in $ d\mu / dT $ is now
\begin{equation} \label{pur1}
  \left. {d (\mu_n-\mu_s) \over dT}  \right |_{T = T_c}
   =  {
         2  (  \partial x  / \partial T  )_s
         +  (  \partial x' / \partial T  )
      \over
         2  (  \partial x  / \partial \mu  )_s
         +  (  \partial x' / \partial \mu  )
      }
    - {
          2  (  \partial x  / \partial T  )_n
          +  (  \partial x' / \partial T  )
      \over
          2  (  \partial x  / \partial \mu )_n
          +  (  \partial x' / \partial \mu )
      }.
\end{equation}

The number of holes in the chain is found as in the usual
  filling of a band,
\begin{equation} \label{xchain}
  x' = 2 \int {dk \over 2\pi} \, n_{\scriptscriptstyle F}^{} (\epsilon_k'),
\end{equation}
  where \( n_{\scriptscriptstyle F}^{} (\epsilon) 
           =  \left[
                   \exp ({\epsilon - \mu  \over T}) + 1
              \right]^{-1} \)
  and $\epsilon_k'$ is the dispersion of the chain band.
Below the transition, there may be superconducting order
  in the chains induced by proximity effect.
However, the gap is expected to be small, especially in the
  vicinity of the transition point and in any case
  it would have no effect on the dependence of $x'$ on~$\mu$.

We will take the chain dispersion in the simple quadratic form
\begin{equation} \label{chdisp}
 \epsilon_k' = \epsilon_0 + \case{1}{2} \beta k^2,
\end{equation}
  where $\epsilon_0$ is the chain band minimum
  relative to the bottom of the plane band
  and $\beta$ is the inverse effective mass.
Then at low temperatures ($T \ll \mu - \epsilon_0$)
\begin{equation} \label{xchainexp}
  x' =  {2^{3/2}  \over \pi \beta^{1/2}}   (\mu - \epsilon_0)^{1/2}
           \left[
                 1 - { \pi^2  \over 24 } {T^2 \over (\mu - \epsilon_0)^2 }
                   + \ldots
           \right].
\end{equation}
It is important to note here that
  the shift $\epsilon_0$ between the plane and chain bands is
  determined by self-consistent distribution of Coulomb charge
  and is a function of~$x'$.
Differentiating (\ref{xchainexp}) we find
\begin{mathletters}
\begin{equation} \label{dx1m}
  \left( 
        {\partial x' \over \partial \mu}
  \right)_T
  =  I_\mu \left(  1 + I_\mu { d \epsilon_0 \over d x' } \right)^{-1},
\qquad
 I_\mu = {1\over \pi}
          \sqrt{2 \over \beta} \,
          {1 \over  \sqrt {\mu - \epsilon_0}}
        = { 4 \over   \pi^2 \beta x'}
\end{equation}
  and
\begin{equation} \label{dx1t}
 \left( 
        {\partial x' \over \partial T}
 \right)_\mu
  = I_T \left(
              1 + I_\mu { d \epsilon_0 \over d x' } 
        \right)^{-1},
\qquad
 I_T = - {\pi \over 6}
          \sqrt{  {2\over \beta} }
          { T \over  (\mu - \epsilon_0)^{3/2}  }
      = - { 16T \over   3 \pi^2 \beta^2 (x')^3}.
\end{equation}
\end{mathletters}
{}From Eqs.~(19), we have the relationship
\begin{equation} \label{dxdx}
  \left( 
        {\partial x' \over \partial T}
  \right)_\mu
  = - { 4 T \over 3 \beta (x')^2 }
     \left(     {\partial x' \over \partial \mu}  \right)_T
\end{equation}
  which allows us to eliminate the unknown dependence $\epsilon_0 (x')$.
To produce the experimental value of $\cusp$,
  we need to take $(\partial x' \!/ \partial \mu )_T = 14$.
For any reasonable value of $\beta$,
  the contribution of $ (\partial x' \!/ \partial T) $ from Eq.~(\ref{dx1t})
  into Eq.~(\ref{pur1}) is negligible (we take $x' = 0.7$).

Now we are in a position
  to estimate the amount of hole transfer between planes and chains.
Using $(\partial x' \!/ \partial \mu )_T = 14$ and Eq.~(\ref{dxdx}), we have
\begin{equation} \label{deltax1}
 \Delta x'
   = x' (T_c) - x'(0)
   = 14 \left(
               \Delta \mu - {2 T_c^2  \over 3 \beta (x')^2}
        \right),
\end{equation}
  where $\Delta \mu = \mu (T_c) - \mu (0)$.
For the in-plane concentration, we have
\begin{mathletters}
\begin{equation}
  \Delta x = 2.45 \Delta \mu - 0.27\, T_c
\end{equation}
  for $\beta_2 = 0$ and
\begin{equation}
  \Delta x = 1.69 \Delta \mu - 0.22\, T_c
\end{equation}
\end{mathletters}
  for $\beta_2 = 0.3$.
Now using the condition $2\, \Delta x + \Delta x' = 0$, we find
\begin{mathletters}
\begin{equation}
  \Delta x = 1.21 { T_c^2 \over \beta (x')^2 } - 0.20\, T_c
\end{equation}
  for $\beta_2 = 0.0$ and
\begin{equation}
  \Delta x = 0.91 { T_c^2 \over \beta (x')^2 } - 0.18\, T_c
\end{equation}
\end{mathletters}
  for $\beta_2 = 0.3$.
For further estimates we will take the average of the two expressions.
The value of $\Delta x$ depends strongly on $\beta$.
$\Delta x$ is positive (holes leave the chains with increasing temperature)
  if $\beta < \beta_{\text{cr}} = 0.45$.
In real units
  (using $J = 0.15$\,eV and the lattice constant $a = 3.8$\,\AA),
  it corresponds to
  effective mass $m^\ast > m^{\ast}_{\text{cr}} = 7.5m_e$.
As an example, for $\beta = 0.20$ ($m^\ast = 17 m_e$)
  the change in the in-plane concentration is $\Delta x = 0.009$,
  so that the in-plane hole concentration at $T_c$ is
  larger by 6\,\% compared to $T = 0$.

\bigskip 
To summarise, we have studied the behaviour of the chemical potential
  near the superconducting phase transition using the
  magnetic fluctuation exchange theory of superconductivity.
We suggested that the found discrepancy with the experimental value
  of the jump in the temperature derivative of the chemical potential
  is be caused by changes in the in-plane hole concentration.
We presented a model which shows that
  even a small hole transfer between chains and planes
  significantly reduces the value of the jump.
Further experiment data (involving other superconductors)
  would lead to better understanding of the problem.

\bigskip 
We are grateful
  to D. van~der~Marel for drawing attention to the problem
  and to D.~I. Khomskii and M.~Yu.\ Kuchiev for stimulating discussions.



\bigskip

\begin{center}  FIGURES  \end{center}

\medskip
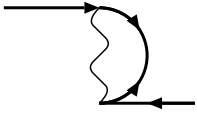
FIG. 1. The diagram for the leading order anomalous self-energy.
\medskip

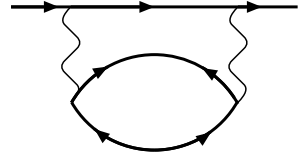
FIG. 2.
  The influence of magnon self-energy on the single electron Green's function.
\medskip

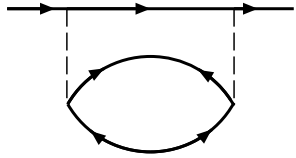
FIG. 3. The influence of the lattice on the single electron Green's function.
\medskip

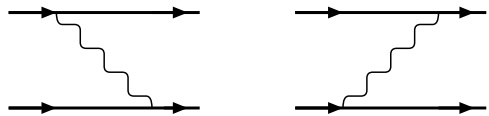
FIG. 4. The diagrams for exchange of one spin wave.
\medskip

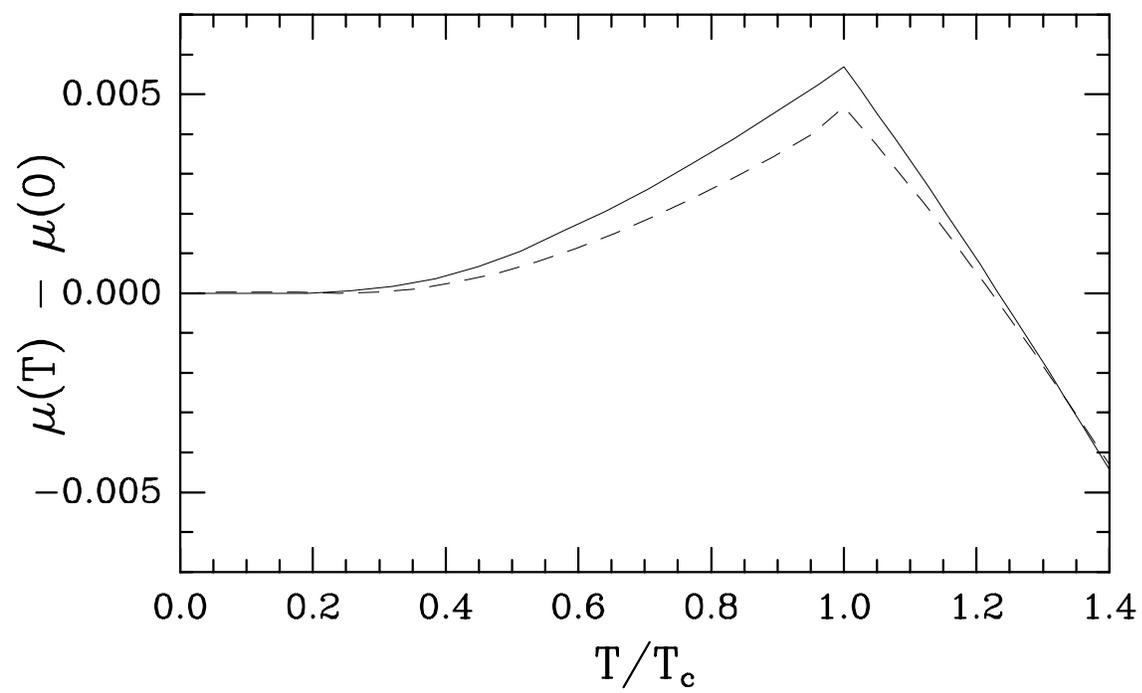
FIG. 5.
  The chemical potential $\mu$ as a function of temperature $T$.
  The dash line is for  $\beta_2 = 0$ 
    and the solid line is for $\beta_2 = 0.3$.
  The hole concentration is constant, $x = 0.15$.

\end{document}